\documentclass[a4paper]{jpconf}
\usepackage{graphicx,amsmath}
\begin{document}
\title{The anatomy of the multilepton anomalies at the LHC and a candidate for a singlet scalar}

\author{Bruce Mellado}

\address{School of Physics and Institute for Collider Particle Physics, University of the Witwatersrand,
Johannesburg, Wits 2050, South Africa}
\address{iThemba LABS, National Research Foundation, PO Box 722, Somerset West 7129, South Africa}

\ead{bmellado@mail.cern.ch}

\begin{abstract}
In this presentation an account of the multi-lepton (electrons and muons) anomalies at the LHC is given. These include the excess production of opposite sign leptons with and without b-quarks, including a corner of the phase-space with a full hadronic jet veto; same sign leptons with and without b-quarks; three leptons with and without b-quarks, including also the presence of a $Z$. Excesses emerge in corners of the phase space where a range of SM processes dominate, indicating that the potential mismodeling of a particular SM process is unlikely to explain them. A procedure is implemented that avoids parameter tuning or scanning the phase-space in order to nullify potential look-else-where effects or selection biases. The internal consistency of these anomalies and their interpretation in the framework of a simplified model are presented. Motivated by the multi-lepton anomalies, a search for narrow resonances with $S\rightarrow\gamma\gamma, Z\gamma$ in association with light jets, $b$-jets or missing transverse energy is performed. The maximum local (global) significance is achieved for $m_S=151.5$\,GeV with 5.1$\sigma$ (4.8$\sigma$).
\end{abstract}

\section{Introduction}

The discovery of a Higgs boson ($h$)~\cite{Higgs:1964ia,Englert:1964et,Higgs:1964pj,Guralnik:1964eu} at the Large Hadron Collider (LHC) by 
ATLAS~\cite{Aad:2012tfa} and CMS~\cite{Chatrchyan:2012ufa} has opened a new chapter in particle physics. Measurements of its properties so far indicate that this $125\,$GeV boson is compatible with those predicted by the Standard Model (SM)~\cite{Chatrchyan:2012jja,Aad:2013xqa}. However, this does not exclude the possible existence of additional scalar bosons as long as the mixing with the SM Higgs is sufficiently small.

One of the implications of a  2HDM+$S$ model, where $S$ is a scalar SM singlet, is the production of multiple-leptons through the decay chain $H\rightarrow Sh,SS$~\cite{vonBuddenbrock:2016rmr}, where $H$ is the heavy CP-even scalar and $h$ is the SM Higgs boson. Excesses in multi-lepton final states were reported in Ref.~\cite{vonBuddenbrock:2017gvy}. In order to further explore results with more data and new final states while avoiding biases and look-else-where effects, the parameters of the model were fixed in 2017 according to Refs.~\cite{vonBuddenbrock:2016rmr,vonBuddenbrock:2017gvy}. This includes setting the scalar masses to $m_H=270$\,GeV, $m_S=150$\,GeV,  treating $S$ as a SM Higgs-like scalar and assuming the dominance of the decays $H\rightarrow Sh,SS$. Excesses in opposite sign di-leptons, same-sign di-leptons, and three leptons, with and without the presence of $b$-tagged hadronic jets were reported in Ref.~\cite{vonBuddenbrock:2019ajh,Hernandez:2019geu}.  Interestingly, the model can explain anomalies in  astro-physics (the positron excess of AMS-02~\cite{PhysRevLett.122.041102} and the excess in gamma-ray fluxes from the galactic centre measured by Fermi-LAT~\cite{Ackermann_2017}) if it is supplemented by a Dark Matter candidate~\cite{Beck:2021xsv} and can be easily extended to account for $g-2$ of the muon~\cite{Sabatta:2019nfg}.

\section{The simplified model}
\label{sec:model}

Here, we succinctly describe the model used to describe the multi-lepton anomalies observed in the LHC data and with which to interpret the above mentioned excesses in astrophysics. The formalism is comprised of a model of fundamental interactions interfaced with a model of cosmic-ray fluxes that emerge from DM annihilation.
The potential for a two Higgs-doublet model with an additional real singlet field $\Phi_S$ (2HDM+$S$) is given as in Ref.~\cite{vonBuddenbrock:2016rmr}:
\begin{equation}
\begin{split}
V(\Phi)
= m^2_{11}\left|\Phi_{1}\right|^2 + m^2_{22}\left|\Phi_{2}\right|^2 - m^2_{12}(\Phi^\dagger_{1}\Phi_{2}+ {\rm h.c.})\notag 
+\frac{\lambda_{1}}{2}\left(\Phi^{\dagger}_{1}\Phi_{1}\right)^2  +\frac{\lambda_{2}}{2}\left(\Phi^{\dagger}_{2}\Phi_{2}\right)^2  \\ +\lambda_{3}\left(\Phi^\dagger_{1}\Phi_{1}\right)\left(\Phi^\dagger_{2}\Phi_{2}\right)\notag
+ \lambda_{4}\left(\Phi^\dagger_{1}\Phi_{2}\right)\left(\Phi^\dagger_{2}\Phi_{1}\right)
+\frac{\lambda_{5}}{2}\Big[\left(\Phi^\dagger_{1}\Phi_{2}\right)^2+ \rm{h.c.}\Big] + \frac{1}{2} m_S^2 \Phi_S^2\notag\\
 + \frac{\lambda_{6}}{8}\Phi^4_{S}  + \frac{\lambda_{7}}{2}\left(\Phi^\dagger_{1}\Phi_{1}\right)\Phi^2_{S} + \frac{\lambda_{8}}{2}\left(\Phi^\dagger_{2}\Phi_{2}\right)\Phi^2_{S}.
\label{pot}
\end{split}
\end{equation}
The fields $\Phi_1$, $\Phi_2$ in the potential are the $SU(2)_L$ Higgs doublets. The first three lines are the contributions of the real 2HDM potential. The terms of the last line are contributions of the singlet field $\Phi_S$. To prevent the tree-level flavour changing neutral currents we consider a ${Z}_2$ symmetry which can be softly broken by the term $m_{12}^2 \neq 0$. After the minimisation of the potential and Electro-Weak symmetry breaking, the scalar sector is populated with three $CP$ even scalars $h, H$ and $S$, one $CP$ odd scalar $A$ and charged scalar $H^\pm$. For more details of this model and associated interactions' Lagrangians and parameter space we refer to  Refs.~\cite{vonBuddenbrock:2016rmr,vonBuddenbrock:2018xar}. Further, we consider interactions of $S$ with three types of DM candidates $\chi_r, \chi_d$ and $\chi_v$ with spins 0, 1/2 and 1, respectively: 

\begin{equation}
\mathcal{L}_{int} =   \frac{1}{2}M_{\chi_{r}}g^{S}_{\chi_{r}}\chi_{r}\chi_{r}S+ {\bar{\chi_{d}}}(g^S_{\chi_{d}}+ig^P_{\chi_{d}}\gamma_{5})\chi_{d}S \notag 
+ g^S_{\chi_{v}}\chi^{\mu}_{v}\chi_{v\mu}S, 
\label{intdm}
\end{equation}

where $g_{\chi_i}$ and $M_{\chi_i}$ are the coupling strengths of DMs with the singlet real scalar $S$ and masses of DM, respectively.

\section{Anatomy of the multilepton anomalies}
\label{sec:anatomy}

We give a succinct description of the different final states and corners of the phase-space that are affected by the anomalies. As discussed in the introduction the anomalies are reasonably well captured by a 2HDM+$S$ model. Here, $H$ is predominantly produced through gluon-gluon fusion and decays mostly into $H\rightarrow SS,Sh$ with a total cross-section in the rage 10-25\,pb~\cite{vonBuddenbrock:2019ajh}. Due to the relative large Yukawa coupling to top quarks needed to achieve the above mentioned direct production cross-section, the  production of $H$ in association with a single top-quark.
These production mechanisms together with the dominance of $H\rightarrow SS,Sh$ over other decays, where $S$ behaves like a SM Higgs-like boson, lead to the a number of final states that can be classified into several groups of final states. Three are the groups of final states where the the excesses are statistically compelling: opposite sign (OS) leptons ($\ell=e,\mu$); same sign (SS) and three leptons ($3\ell$) in association with $b$-quarks; SS and $3\ell$ without $b$-quarks. 
In the sections below a brief description of the final states is given with emphasis on the emergence of new excesses in addition to those reported in Refs.~\cite{vonBuddenbrock:2017gvy,vonBuddenbrock:2019ajh,Hernandez:2019geu}, when appropriate. The new excesses reported here are not the result of scanning the phase-space, but the result of looking at pre-defined final states and corners of the phase-space, as predicted by the model described above.  

\begin{table}[t]
\begin{center}
      \begin{tabular}{c|c|c|c}
      \hline\hline
Final state & Characteristics & SM backgrounds & Significance \\
\hline
$\ell^+\ell^-$+$b$-jets & $m_{\ell\ell}<100$\,GeV, low $b$-jet mult. & $t\overline{t}, Wt$ & $>5\sigma$ \\ 
$\ell^+\ell^-$+jet veto & $m_{\ell\ell}<100$\,GeV & $W^+W^-$ & $\approx 3\sigma$ \\ 
$\ell^\pm\ell^\pm, 3\ell$ + $b$-jets & Moderate $H_T$ & $t\overline{t}W^{\pm}, t\overline{t}t\overline{t}$ & $>3\sigma$ \\ 
$\ell^\pm\ell^\pm, 3\ell, n_b=0$ & In association with $h$ & $W^{\pm}h, WWW$ & $\approx 4.5\sigma$ \\ 
$Z(\rightarrow\ell\ell)\ell, n_b=0$ & $p_{TZ}<100$\,GeV & $ZW^{\pm}$ & $> 3\sigma$ \\ 
    \hline
      \end{tabular}
      \caption{Summary of the status of the multi-lepton anomalies at the LHC, where $\ell=e,\mu$.}
      \label{tab:anatomymultilepton}
      \end{center}
\end{table}

\subsection{Opposite sign di-leptons}
\label{sec:OS}

The production chain $pp\rightarrow H\rightarrow SS,Sh\rightarrow\ell^+\ell^-+X$,  is the most copiously multi-lepton final state. Using the benchmark parameter space in Ref.~\cite{vonBuddenbrock:2018xar}, the dominant of the singlet are $S\rightarrow W^+W^-,b\overline{b}$. This will lead to OS leptons with without $b$-quarks. The most salient characteristics of the final states are such that the di-lepton invariant mass $m_{\ell\ell}<100$\,GeV where the bulk of the signal is produced with low $b$-jet multiplicity, $n_b<2$~\cite{vonBuddenbrock:2019ajh}. The dominant SM background in events with $b$-jets is $t\overline{t}+Wt$. The $b$-jet and light-quark of the signal is significantly different from that of top-quark related production mechanisms. As a matter of fact, excesses are seen when applying a full jet veto, top-quark backgrounds become suppressed and where the dominant backgrounds is non-resonant $W^+W^-$ production~\cite{vonBuddenbrock:2017gvy,ATLAS:2019rob,vonBuddenbrock:2019daj}.\footnote{
The CMS experiment presented the comparison of the yields in the data to a MC~\cite{CMS:2020mxy}. The MC does not describe simultaneously the di-lepton invariant mass and transverse momentum. As such, for the purposes of this study this data set is inconclusive, where we encourage the CMS experiment to provide differential measurements similar to those performed by ATLAS in Ref.~\cite{ATLAS:2019rob}.} A review of the NLO and EW corrections to the relevant processes can be found in Refs.~\cite{vonBuddenbrock:2019ajh,vonBuddenbrock:2019daj}, where to date the $m_{\ell\ell}$ spectra at low masses remains unexplained by MC tools. A measurement of the differential distributions in OS events with $b$-jets with Run 2 data further corroborates the inability of current MC tools to describe the $m_{\ell\ell}$ distribution~\cite{ATLAS:2019hau}. A summary of  deviations for this class of excesses is given in Tab.~\ref{tab:anatomymultilepton}.

\subsection{SS and $3\ell$ with $b$-quarks}
\label{sec:SS3lb}

The associated production of $H$ with top quarks lead to the anomalous production of SS and $3\ell$ in association with $b$-quarks with moderate scalar sum of leptons and jets, $H_T$. The elevated $t\overline{t}W^{\pm}$ cross-section measured by the ATLAS and CMS experiments can be accommodated by the above mentioned model~\cite{vonBuddenbrock:2019ajh,vonBuddenbrock:2020ter}. Based on a number of excesses involving $Z$ bosons, in Ref.~\cite{vonBuddenbrock:2018xar} it was suggested that the $CP$-odd scalar of the 2HDM+$S$ model could be as heavy as $m_A\approx500$\,GeV, where the two leading decays would be $A\rightarrow t\overline{t},ZH$. The cross-section for the associated production $pp\rightarrow t\overline{t}A$ with $A\rightarrow t\overline{t}$ would correspond to $\approx 10$\,fb. This is consistent with the elevated $t\overline{t}t\overline{t}$ cross-section reported by ATLAS and CMS~\cite{CMS:2019rvj,ATLAS:2020hpj,ATLAS:2021kqb}. The combined significance of the excesses related to the cross-section measurements of $t\overline{t}W^{\pm}$ and $t\overline{t}t\overline{t}$ surpass 3$\sigma$, as detailed in  Tab.~\ref{tab:anatomymultilepton}. The ATLAS collaboration has reported a small excess in the production of four leptons with a same flavor OS pair consistent with a $Z$ boson, where the four-lepton invariant mass, $m_{4\ell}<400$\,GeV~\cite{ATLAS:2021wob}. This excess can also be accommodated by the production of $A\rightarrow ZH$.

\subsection{SS and $3\ell$ without $b$-quarks}
\label{sec:SS3lnob}

The production chain $pp\rightarrow H\rightarrow SS,Sh$ can give rise to SS and $3\ell$ events, where $b$-jet activity would be depleted compared to production mechanism considered in Sec.~\ref{sec:SS3lb}. The potential impact on the measurement of the production of the SM Higgs boson in association with a $W$ boson and other measurements in the context discussed here was reported in Ref.~\cite{Fang:2017tmh}. A survey of available measurements of the signal yield of the $Wh$ production was performed in Ref.~\cite{Hernandez:2019geu}. A deviation of 3.8$\sigma$ with respect to the $Wh$ yield in the SM in corners of the phase-space predicted by the simplified model. The CMS experiment has recently reported the signal strength of the $Vh, V=Z,W^{\pm}$ production with the $h\rightarrow W^+W^-$ decay for low and high $V$ transverse momentum~\cite{CMS:2021ixs}. The signal strength for $Vh$ with the $V$ transverse momentum, $p_{TV}<150$\,GeV, where the BSM signal is concentrated, is $2.65^{+0.69}_{-0.64}$. This deviates from the SM value by an additional 2.6$\sigma$. In order to reconcile observed excesses in Secs.~\ref{sec:OS} and~\ref{sec:SS3lb} with the ones described here, it is necessary to assume the dominance of the $H\rightarrow SS$ decay over $H\rightarrow Sh$~\cite{Hernandez:2019geu}. Another important prediction of the simplified model is the elevated $WWW$ cross-section. The ATLAS experiment reports a signal strength of $1.66\pm0.28$~\cite{ATLAS-CONF-2021-039}.\footnote{CMS~\cite{CMS:2020hjs} pursues a different approach compared to the more inclusive selection performed by ATLAS. For instance, the requirement that the azimuthal separation between the vector of the three leptons and the missing transverse energy be greater than 2.5\,rad and other requirements suppress the contribution from the BSM signal considered here.} The latter includes the $Wh\rightarrow WWW^*$ production, hence it is not added to the combination due to partial double counting. Another final state of interest is the production of $ZW^{\pm}$ events where $Z$ transverse momentum, $p_{TZ}<100$\,GeV with depleted $b$-jet activity. Excesses were reported in Ref.~\cite{vonBuddenbrock:2019ajh}. The CMS experiment has recently reported an important excess in events with $3\ell$ in association with one and two jets used for the measurement of $Zh, h\rightarrow W^+W^-$ production, where $ZW^{\pm}$ is the dominant background~\cite{CMS:2021ixs}. As the analysis of the excess in the context of the simplified model described here is in progress, the significance of this excess is not added to the combination reported in Tab.~\ref{tab:anatomymultilepton}. 

\section{Candidate of Singlet Scalar}
\label{sec:candidate}

\begin{figure}[t]
  \centering
   \includegraphics[width=0.40\textwidth]{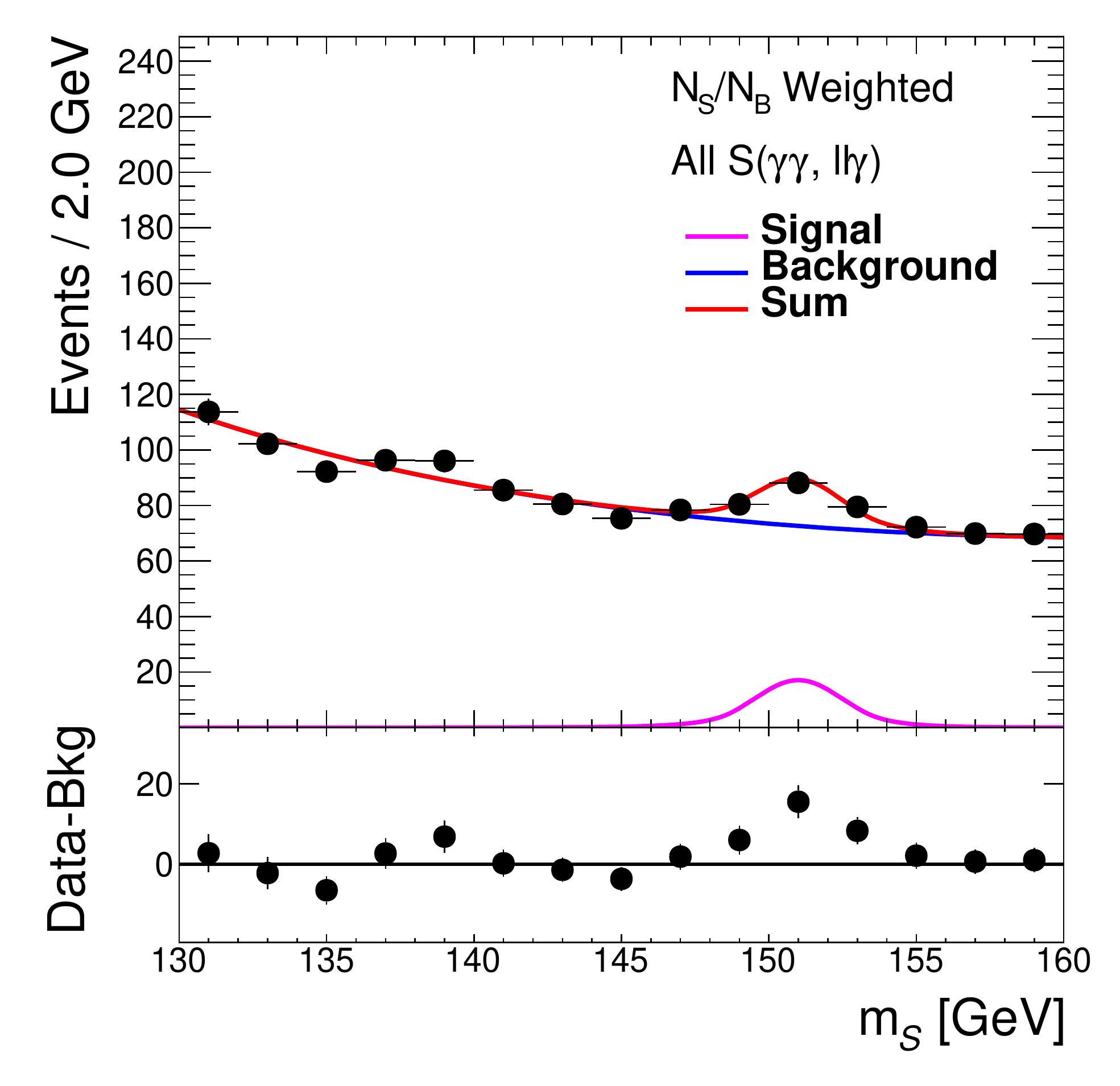}
  \includegraphics[width=0.55\textwidth]{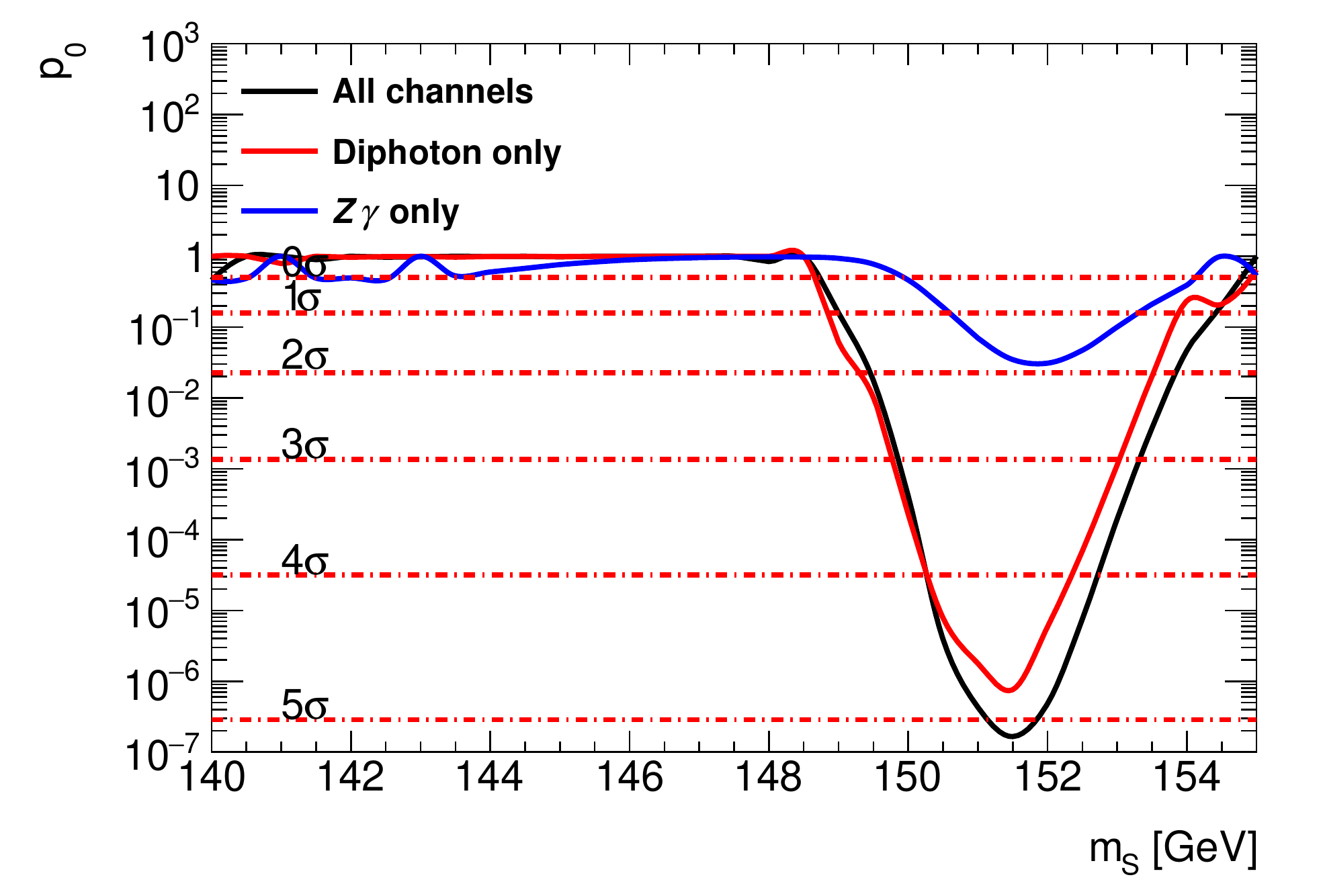}
            \caption{
        Combination (left) of fits to all the categories in the $\gamma\gamma$ and $Z\gamma$ spectra (see text). The data points are S/B weighted. The combined local $p$-value (right) as a function of $m_{S}$ using inputs  (see text).}
    \label{fig:invmass2}
 \end{figure}

The multi-lepton anomalies described here seem to be relatively well accommodated by 2HDM+$S$ model with a sizeable direct production of $H\rightarrow SS,Sh$. This motivates the search for narrow resonances pertaining to $S\rightarrow\gamma\gamma, Z\gamma$ in association with light jets, $b$-jets or missing transverse energy. A survey of all publicly available data is performed~\cite{Crivellin:2021ubm} in the mass range motivated by the multi-lepton anomalies, in particular the di-lepton invariant mass spectra of the excesses. The region of interest is in the range $130<m_S<160$\,GeV. Figure~\ref{fig:invmass2} displays the signal to background weighted combination of the mass spectra, where the red curve displays the combined fit. The signal normalisation in each of the categories is allowed to float. Figure~\ref{fig:invmass2} (right) shows the local $p$-value distribution  in the range 140-155\,GeV. The upper and lower bounds of the search are determined by the availability of data reported by the experiments. The lowest local $p$-value is achieved for $m_S=151.5$\,GeV corresponding to 5.1$\sigma$. Taking into account the look-else-where effect the global significance in the range 140-155\,GeV is 4.8$\sigma$. Searches for $H\rightarrow SS^{(*)},Sh\rightarrow \gamma\gamma b\overline{b},\tau^+\tau^-b\overline{b}$ in asymmetric configurations, not performed  by the experiments before, are well motivated~\cite{Crivellin:2021ubm}.

Interestingly, as the LEP experiments reported a mild excess (2.3$\sigma$) in the search for a scalar boson ($S^\prime$)~\cite{LEPWorkingGroupforHiggsbosonsearches:2003ing} using the process $e^+e^-\rightarrow Zh(\rightarrow b\overline{b})$ at 98\,GeV for the invariant $b\overline{b}$ mass, asymmetric  $\gamma\gamma b\overline{b}$ final states could also originate from the decay $H\to SS^\prime$. This is further supported by the CMS result reporting similar excesses with Run 1 data and 35.9\,fb$^{-1}$ of Run 2 data~\cite{CMS:2018cyk}, with a local significance of 2.8$\sigma$ at 95.3\,GeV. In this context searches for $H\to S^{(\prime)}(\to \gamma\gamma,b\bar b)S(\to {\rm invisible})$ are well motivated.
 
\section{Conclusions}
\label{sec:conclusions}

This proceedings provide an update of the multi-lepton anomalies at the LHC since Ref.~\cite{vonBuddenbrock:2019ajh}, where new data and final states and corner of the phase-space predicted by the 2HDM+$S$ model  display excesses with respect to the SM. A summary of the multi-lepton anomalies is given in Tab.~\ref{tab:anatomymultilepton}. Motivated by the fact that the $H\rightarrow SS,Sh$ decay describes reasonably well the multi-lepton anomalies a search for narrow resonances with $S\rightarrow\gamma\gamma, Z\gamma$ in association with light jets, $b$-jets or missing transverse energy is performed. The maximum local (global) significance is achieved for $m_S=151.5$\,GeV with 5.1$\sigma$ (4.8$\sigma$).

\section*{References}
\bibliographystyle{iopart-num}
\bibliography{multilepton}

\end{document}